\documentclass[10pt,conference,compsocconf]{IEEEtran}
\usepackage{times}
\usepackage{caption}
\usepackage{tikz}
\usetikzlibrary{er,positioning, shapes, shapes.geometric, arrows, trees, shapes.multipart, calc}
\tikzstyle{process} = [rectangle, text width=10em,minimum height = .9cm,minimum width=3cm,text centered, draw=black, fill=white]
\tikzstyle{titlef} = [draw=none, text width=10em,minimum height = .9cm,minimum width=3cm,text centered, font=\bfseries]

\tikzstyle{arrow} = [thick,->,>=stealth]
\tikzstyle{someline} = [thick,-,>=stealth]

\newcommand{\itsecn}{IT Security}
\newcommand{\itsec}{IT Security }

\newcommand{\flexn}[1]{``#1''}
\newcommand{\flex}[1]{\flexn{#1}}
\newcommand{\enn}[1]{{\em #1}}
\newcommand{\en}[1]{\enn{#1} }

\newcommand{\stagn}{\enn{S-Tag}}
\newcommand{\stag}{\en{S-Tag}}

\newcommand{\mtagn}{\enn{S-Mark}}
\newcommand{\mtag}{\en{S-Mark}}

\newcommand{\sscrum}{{\em Secure Scrum }}

\begin{document}
\pagenumbering{gobble}
%
% paper title
% can use linebreaks \\ within to get better formatting as desired
% \title{\textbf{Development of Secure Software Using Secure Scrum\\[0.2ex]}}
\title{\textbf{Secure Scrum: Development of Secure Software with Scrum\\[0.2ex]}}
% author names and affiliations
% use a multiple column layout for up to three different
% affiliations
\author{
\IEEEauthorblockN{~\\[-0.4ex]\large Christoph Pohl\\ and Hans-Joachim Hof\\[0.3ex]\normalsize}
\IEEEauthorblockA{MuSe - Munich IT Security Research Group\\
Munich University of Applied Sciences\\
Email: {\tt christoph.pohl0@hm.edu, hof@hm.edu}}
}

\maketitle

\begin{abstract}
%\boldmath
Nowadays, the use of agile software development methods like Scrum is common in industry and academia. Considering the current attacking landscape, it is clear that developing secure software should be a main concern in all software development projects. 
In traditional software projects, security issues require detailed planning in an initial planning phase, typically resulting in a detailed security analysis (e.g., threat and risk analysis), a security architecture, and instructions for security implementation (e.g., specification of key sizes and cryptographic algorithms to use). 
Agile software development methods like Scrum are known for reducing the initial planning phases (e.g., sprint 0 in Scrum) and for focusing more on producing running code. 
Scrum is also known for allowing fast adaption of the emerging software to changes of customer wishes. For security, this means that it is likely that there are no detailed security architecture or security implementation instructions from the start of the project. 
It also means that a lot of design decisions will be made during the runtime of the project. 
Hence, to address security in Scrum, it is necessary to consider security issues throughout the whole software development process. Secure Scrum is a variation of the Scrum framework with special focus on the development of secure software throughout the whole software development process. 
It puts emphasis on implementation of security related issues without the need of changing the underlying Scrum process or influencing team dynamics. 
Secure Scrum allows even non-security experts to spot security issues, to implement security features, and to verify implementations.  
A field test of Secure Scrum shows that the security level of software developed using Secure Scrum is higher then the security level of software developed using standard Scrum.
\end{abstract}


% For peer review papers, you can put extra information on the cover
% page as needed:
% \ifCLASSOPTIONpeerreview
% \begin{center} \bfseries EDICS Category: 3-BBND \end{center}
% \fi
%
% For peer review papers, this IEEEtran command inserts a page break and
% creates the second title. It will be ignored for other modes.
\IEEEpeerreviewmaketitle
\section{Introduction}
Nowadays, software is all around us, even refrigerators now have network support and run a whole bunch of software. 
As software is so ubiquitous today, software bugs that lead to successful attacks on software systems are becoming a major hassle. 
Hence, modern software development should focus on SECURE software. 
At the moment, Scrum \cite{agile_sd_manifesto_2014}\cite{schwaber_scrum_1997} is a very popular software development framework. This paper presents Secure Scrum, an extension of the Scrum framework that helps developers, even non-security experts, to develop secure software. 

Scrum groups developer in small developer team that have a certain autonomy to develop software. It is assumed that all developers can implement all tasks at hand. Software is incrementally developed in so called sprints. A sprint is a fixed period of time (between 2 and 4 weeks). During a sprint, the team develops an increment of the current software version, typically including a defined number of new features or functionality, which are described as user stories. User stories are used in Scrum to document requirements for a software project. All user stories are stored in the Product Backlog. During the planning of a sprint, user stories from the Product Backlog are divided into tasks. These tasks are stored in the Sprint Backlog.
A so called Product Owner is the single point of communication between customer and developer team. The Product Owner also prioritizes the features to implement. Standard Scrum does not include any security-specific parts. 

One major driver of software security in Secure Scrum is the identification of security relevant parts of a software project. The security relevance is then made visible to all team members at all times. This approach is considered to increase the security level because developers place their focus on things that they had evaluated themselves, which they fully understand, and when their prioritization of requirements does not differ from prioritization of others \cite{riemenschneider_explaining_2002}\cite{vijayasarathy_drivers_2012}.

Secure Scrum aims on achieving an appropriate security level for a given software project. The term "appropriate" was chosen to avoid costly over engineering of IT security in software projects. The definition of an appropriate security level is the crucial point in resource efficient software development (e.g., time and money are important resources during software development). 
For the definition of an appropriate security level, Secure Scrum relies on the definition in \cite{Herley:2014ke}: Software needs to be secured until it is no longer profitable for an intruder to find and exploit a vulnerability. This means that an appropriate security level is reached once the cost to exploit a vulnerability is higher then the expected gain of the exploit. So, Secure Scrum offers a way to not only identify security relevant parts of the project but to also judge on the attractiveness of attack vectors in the sense of ease of exploitation.

Related to the identification of security issues, the developers need to implement features to avoid these potential security risks. 
In Scrum, each team member is responsible for the completeness of his solution (Definition of Done). However, there is a huge number of choices of methodologies to verify completeness. 
%Thus, the toolset must be able to integrate different verification methods. 
%This leads to the issue that this approach needs to help the team member with verification, but without the use of predefined verification methods.
This means that a team member can use any method for verification (same as with normal tests, Scrum does not tell the developer how to test). Secure Scrum helps developers to identify appropriate security testing means for security relevant parts of a software project.

One last challenge solved by Secure Scrum is the availability of know how when needed. %In Scrum, each team member is responsible for his work, this also means that the team member needs the knowledge to solve the requested task. 
%Nowadays, the availability of security know how and experience does not reflect the importance of this issue. 
Secure Scrum assumes that the vast majority of requirements should and could be handled by the team itself to keep many benefits of Scrum. 
However, for some security related issues, it could be necessary or more cost effective to include external resources like security consultants in the project.
Secure Scrum offers a way to include these external resources into the project without breaking the characteristics of Scrum and with little overhead in administration.
%To keep it flexible Secure Scrum does not define any level of knowledge related to \itsec, it only defines that the team should be able to solve most of the challenges (including the necessary \itsec parts) on their own.

The rest of this paper is structured as follows:
The following section summarizes related work. Section \ref{securescrum} shows the design of Secure Scrum in detail. Secure Scrum is evaluated in a field test in Section \ref{eval}. Section \ref{sec:conclusion} summarizes the findings of this paper.

\section{Related Work}\label{relwork}
There are several methods for achieving software security, e.g., Clean Room \cite{Mills:1991ul}, Construction by Correctness \cite{Hall:2002ht}, CMMI-DEV \cite{Chrissis:2011vp}\cite{Glazer:2008up}, etc.
However, these methods cannot be used in Scrum as they clash with the characteristics of agile software development and specifically Scrum. 
Construction by Correctness \cite{Hall:2002ht}  for example, advocate formal development in planning, verification and testing.
This is completely different to agility and flexible approaches like agile methodologies. 
Other models like CMMI-DEV \cite{Chrissis:2011vp}\cite{Glazer:2008up} can deal with agile methods, but they are process models.
The main difference is that CMMI focuses on processes and agile development on the developers \cite{Glazer:2008up}.
This means that Scrum and other agile methodologies are developer centric, while CMMI is more process oriented.
Concepts like Microsoft SDL \cite{howard_security_2009} are designed to integrate agile methodologies, but is also self-contained. 
It can not be plugged into Scrum or any other agile methodology. 
Scrum focuses on rich communication, self-organisation, and collaboration between the involved project members. This conflicts with formalistic and rigid concepts.

To sum it up, the major challenge of addressing software security in Scrum is not to conflict with the agility aspect of Scrum.

S-Scrum \cite{mougouei_s-scrum:_2013} is a \flex{security enhanced version of Scrum}.
It modifies the Scrum process by inserting so called spikes.
A spike contains analysis, design and verification related to security concerns.
Further, requirements engineering (RE) in story gathering takes effect on this process.
For this, the authors describe to use tools like Misuse Stories \cite{sindre_eliciting_2005}.
This approach is very formalistic and needs lot of changes to standard Scrum, hence hinders deployment in environments already using Scrum. Secure Scrum in contrast does not change standard Scrum.

Another approach is described in \cite{azham_security_2011}.
It introduces a Security Backlog beside the Product Backlog and Sprint Backlog.
Together with this artifact, they introduce a new role.
The security master should be responsible for this new Backlog.
This approach introduces an expert, describes the security aware parts in the backlog, and is adapted to the Scrum process. However, it lacks flexibility (as described in the introduction) and does not fit naturally in a grown Scrum team. Also, the introduction of a new role changes the management of the project. With this approach it is not possible to interconnect standard Scrum user stories with the introduced security related stories. Secure Scrum in contrast keeps the connect between security issues and user stories of the Product Backlog respectively tasks of the Sprint Backlog.

%Keramati introduces in \cite{keramati_integrating_2008} an algorithm which should define the effort to perform security related tasks within agile methodologies.
%While this is an interesting approach, it does not give hints how to include security aware parts in Scrum.
%It does not describe how to implement \itsec within a Scrum process.

In \cite{risk_protection_2014} an informal game (Protection Poker) is used to estimate security risks to explain security requirements to the developer team.
The related case study shows that this is a possible way to integrate security awareness into Scrum.
It solves the problem of requirements engineering with focus on \itsecn.
However, it does not provide a solution for the implementation and verification phase of software development, hence it is incomplete. Secure Scrum in contrast provides a solution for all phases of software development.

Another approach is discussed in \cite{bostrom_extending_2006}.
An XP Team is accompanied by a security engineer.
This should help to identify critical parts in the development process.
Results are documented using abuse stories.
This is similar to the definition in \cite{peeters_agile_2005}.
This approach is suitable for XP-Teams but not for Scrum. 

%A very formalistic approach is the Microsoft SDL \cite{howard_security_2009}.
%This attempt is based on the classical spiral model with the ability to be more Agile friendly \cite{baca_agile_2011}. 
%Hence, this approach integrates security and uses agile methodologies but it is also a self contained method.
%Thus, it is a self contained method which does not the solve the problem to integrate security into Scrum. \bfn{Verstehe ich nicht. Auch schon oben erwähnt, einmal streichen!}

To sum it up, none of the related work mentioned above integrates well into Scrum, allows for easy adaption for teams already using standard Scrum, and focuses on all phases of software development. Secure Scrum in contrast solves all of these problems. The design of Secure Scrum is described in detail in the following.

\section{Design of Secure Scrum}
\label{securescrum}

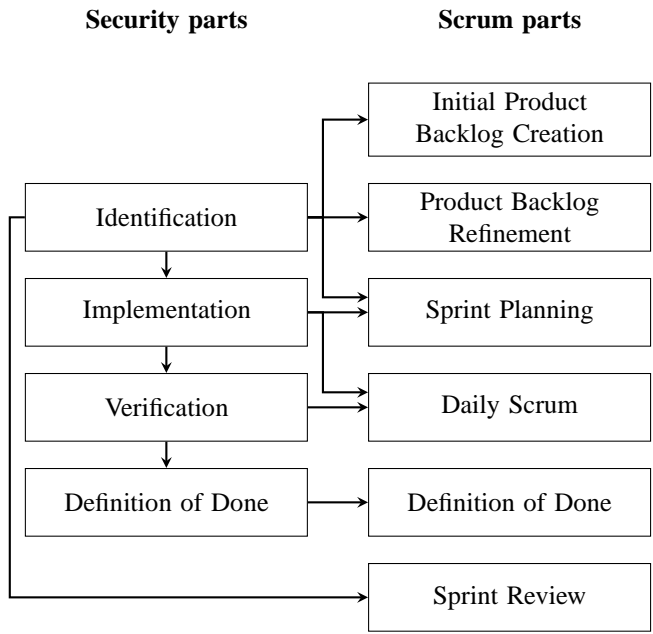
\begin{figure}[!htpb]
\centering%
%\fbox{
\begin{tikzpicture}
%\node (identification) [process] {Identification};
\node (scrum) [titlef] {Scrum parts};
\node (sss) [titlef, left=.8cm of scrum] {Security parts};

\node (productback) [process,below=.35cm of scrum] {Initial Product Backlog Creation};
\node (productbackref) [process, below=.35cm of productback] {Product Backlog Refinement};
\node (identification) [process, left=.8cm of productbackref] {Identification};
\node (sprintplan) [process, below=.35cm of productbackref] {Sprint Planning};

\node (sprintbacklog) [process, below=.35cm of sprintplan] {Daily Scrum};
\node (dod2) [process, below=.35cm of sprintbacklog] {Definition of Done};
\node (sprintreview) [process, below=.35cm of dod2] {Sprint Review};

\node (implementation) [process, left=.8cm of sprintplan] {Implementation};
\node (verification) [process, below=.35cm of implementation] {Verification};
\node (dod) [process, below=.35cm of verification] {Definition of Done};

%\draw [->] (client.south)-|(revprox.north);
\draw [arrow](identification) -- (implementation); 
\draw [arrow](implementation) -- (verification); 
\draw [arrow](verification) -- (dod); 

\draw [arrow](identification.east) --++ (.2,0) |- (productback.west);
\draw [arrow](identification.east) -- (productbackref.west);
\draw [arrow](identification.east) --++ (.2,0) |- ([yshift=.2cm]sprintplan.west);

\draw [arrow](implementation.east) -- (sprintplan.west);
\draw [arrow](implementation.east) --++ (.2,0) |- ([yshift=.2cm]sprintbacklog.west);

\draw [arrow](verification.east) -- (sprintbacklog.west);
\draw [arrow](dod.east) -- (dod2.west);

\draw [arrow](identification.west) --++ (-.2,0) |- (sprintreview.west);
\end{tikzpicture}%} 
\caption{Integration of Secure Scrum components into standard Scrum} 
\label{fig:overview}
\end{figure}

Secure Scrum consists of four components:
\begin{itemize}
\item Identification component
\item Implementation component
\item Verification component
\item Definition of Done component
\end{itemize}
These four components are put on top of the standard Scrum framework. Secure Scrum influences six stages of the standard Scrum process as can be seen in Figure \ref{fig:overview}.

The identification component is used to identify security issues during software development. Security issues are marked in the Product Backlog of Scrum. The identification component is used during the initial creation of the Product Backlog as well as during Product Backlog Refinement, Sprint Planning, and Sprint Review.

The implementation component raises the awareness of the Scrum team for security issues during a sprint. The implementation component is used in Sprint Planning, as well as during the Daily Scrum meetings.

The verification component ensures that team members are able to test the software with the focus on \itsecn. The verification component gets managed within the Daily Scrum meeting.

The Definition of Done component enables the developers to define the Definition of Done for security related issues as postulated in standard Scrum.

These four components of Secure Scrum are described in detail in the following subsections.

\subsection{Identification Component}\label{sec:ident}
The identification component is used to identify and mark security-relevant user stories. Secure Scrum takes a value-oriented approach to security as described in the Introduction. It focuses security implementation effort on parts of the emerging software that are of high value for the stakeholders.
The identification component of Secure Scrum is used during initial Product Backlog creation, during Sprint Planning, as well as during Product Backlog Refinement.

In a first step, stakeholders (may be represented by the Product Owner) and team members rank the different user stories according to their loss value. 
The loss value of a user story is not the cost of development neither the benefit of the functionality that implements the user story. The loss value of a user story is the loss that may occur whenever the functionality that implements the user story gets attacked or data processed by this functionality gets stolen or manipulated. For example one can formulate \flex{Whenever someone will get access to these data, our company will have high damage}. 
Even better the cost gets listed with a numerable value like USD or Euro. 

In a next step, stakeholders and team members evaluate misuse cases and rank them by their risk. 

At this point, it can be useful to incorporate external security expertise to moderate by asking the right questions and proposing security aware user stories.

\begin{figure}[!htpb]
\centering%
%\fbox{
\begin{tikzpicture}
\node (userstory1) [process] {User Story A\\Story$\dots$};
\node (userstory2) [process, below of=userstory1] {User Story B\\Story$\dots$};
\node (userstory3) [process, below of=userstory2] {User Story C\\Story$\dots$};

\node (stag1) [process, right=1cm of userstory2] {S-Tag A\\Story (Description)$\dots$};
\node (stag2) [process, below of=stag1] {S-Tag B\\Story (Description)$\dots$};

\node (d1) [above right=.8cm and .3cm of userstory2] {S-Mark};
\node (d2) [above right=.3cm and .9cm of userstory2] {Connection};
\node (d3) [above right=.3cm and -1.5cm of stag1] {Description};

\node (m1) [rectangle, fill=black, minimum width=.2cm, minimum height=.2cm, right=-.3cm of userstory2]{};

\node (m2) [rectangle, fill=black, minimum width=.2cm, minimum height=.2cm, right=-.3cm of userstory3]{};
%\draw [someline](userstory1.east) --++ (.2,0) |- ([yshift=.2cm]stag1.west); 
\draw [someline](userstory2.east) --++ (.2,0) |- (stag1.west);

\draw [someline]([yshift=.2cm]userstory3.east) --++ (.2,0) |- ([yshift=-.2cm]stag1.west);
\draw [someline](userstory3.east) --++ (.2,0) |- (stag2.west);

\draw [-](d1.west) -- (m1.north);
\draw [-](d2.west) -- ($ (userstory2.east) !.5! (stag1.west) $);
\draw [-](d3.west) -- (stag1.north);

\end{tikzpicture}%} 
\caption{Usage of S-Tags to mark user stories in the Product Backlog and to connect user stories to descriptions of security related issues.} 
\label{fig:stag}
\end{figure}
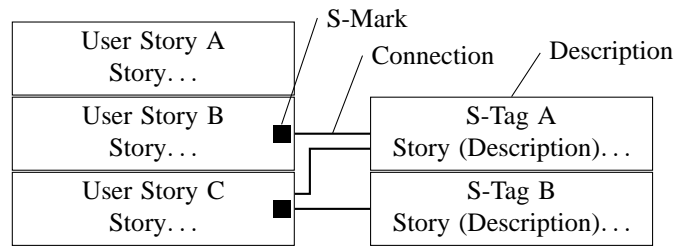

After finalization of the identification component, team members and stakeholders have  a common understanding of security risks in the Product Backlog. To document this understanding in the Product Backlog, Secure Scrum uses so called \stagn s. 
Figure \ref{fig:stag} shows the basic principle of an \stagn. An \stag consists of one or more \mtagn s, a Backlog artifact, and a connection between the Backlog artifact and one or more \stagn s. An \stag identifies Product Backlog items that have security relevance with a Marker called \mtagn. This ensures that security sensitive items in the Product Backlog are visible at all times. The technology behind the \mtag is negligible (it can be a red background, a dot, or something else), it only must ensure that a Product Backlog item with security relevance contrasts to other Backlog items.

An \stag describes a security concern. A detailed description of the security issue helps the Scrum team to understand the security concern. The description of the security concern itself can be formulated in a separate Backlog item. This can be a user story, misuse story, abuse story, or whatever a team decides to use as description technology. The description may include elements from a knowledge base that gives advice on how to deal with this specific security concern. If such a knowledge base is maintained over the course of several projects, it is very likely a valuable source of information for the Scrum team.

An \stag links one security concern to one or more Backlog items. A security concern is any security related problem, attack vector, task, or security principle that should be considered during implementation. One-to-many-connections between security concern and affected Product Backlog items allow for grouping of items that share the same security concern (and hopefully may use the same security mechanisms) as well as expressing security on a high level. To express the connections, unique IDs can be used.

%Whenever an artifact gets connected with the problem description, it is marked with the \mtag.

%However a \stag can be self descriptive too.
%Thenever a misuse story (or any other description technology) does not affect any other artifact tha \mtag is used on itself and connects to itself.

\subsection{Implementation Component}\label{sec:implement}
The Scrum framework has a focus on implementation. Thus, during implementation every team member needs to be aware of the top priority topics of the project.
This means that most of the requirements (functionalities) are described in the Product Backlog. This includes the \stagn s. To ensure that security concerns are visible in daily work, they must be present in the Sprint Backlog. 

Usually, a sprint implements a subset of functionalities (for example user stories).
During a sprint, some user stories are broke down to tasks (or similar conceptual parts). Whenever a user story is marked with an \mtagn, the corresponding \stag must also be present in the correspondding sprint. An \stag can be handled like any other Backlog item. But whenever an \stag gets splitted into tasks, the tasks must also be marked with an \mtag and connected to the original \stagn. This ensures that developers are always aware of the original security concern and it can be linked back to the origin description. 

This approach determines that the interconnection will enhance the awareness of the developer for the security problems.

\subsection{Verification Component and Definition of Done Component}\label{verif}
Not only \stagn s help developers to be aware of security relevance of  user stories, they can also be used to identify requirement for the verification of the emerging software. In the first place, \stagn s clearly identify parts of the emerging software that need security verification. In the second place, \stagn s are useful to estimate the effort for verification.

\sscrum proposes two different approaches for verification and therefore the Definition of Done. For further simplification, the term ``task'' is used for some work that is performed by one developer in one sprint and that needs one Definition of Done. Whenever the verification process (whatever the developer or team chooses to use) for one task can be performed during the same sprint and from the same developer, the verification must be part of the task. This ensures that the verification must be part of the Definition of Done. However, it is possible that a developer does not have the required knowledge for verification, or the verification needs external resources, extra time for testing, or anything else that hinders an immediate verification. In this case, the verification cannot be part of the Definition of Done. In such cases, a new task must be created which inherits only the verification part. This new task must be marked with an \mtag and should be connected to the original \stagn, together with the original task. Then, the developer can define the Definition of Done without the verification, hence a Definition of Done compatible to standard Scrum is available.

The proposed approach for the definition of the Definition of Done ensures that the connection between an \mtag and its corresponding \stag keeps existing throughout the project, hence no security concern can get lost or stay untested.

\subsection{Integration of External Ressources}\label{sec:integrate}
IT security knowledge may be rare in a Scrum team or special knowledge not present in the Scrum team may be necessary for certain parts of the emerging software (e.g., implementation and testing of cryptographic algorithms). Secure Scrum offers ways to include external resources (e.g., security consultants) in all components of Secure Scrum. External resources could have one or more of the following three functions:
\begin{itemize}
\item Enhance knowledge
\item Solve challenges 
\item Provide external view
\end{itemize}
These three functions are described in the following.

Enhancing knowledge: This function includes security-related training for the Scrum team to help them to gain a better understanding of a specific security-related area. Doing so on the job during a project offers a chance to teach IT security with a specific example at hand (e.g., a certain \stag that is linked to many user stories) and may be more efficient than security training during two projects. Training may be necessary for aspects that are not part of everyday work, e.g., the usability of security mechanisms \cite{Hof2013,Hof2012}.

Solving Challenges: Some \stagn s represent hard security challenges that require special expertise or special experience, such that it is more cost efficient to let external resources solve this challenge.  To avoid breaches in Scrum, it is necessary that these external solutions can be handled like a tool, a well defined part of development, a framework, or a ``black box'', which is ready to use. This means that this external solution should be encapsulated and therefore does not influence Scrum or the Scrum team. For example, this can be a functional part of software (with special \itsec concerns) or parts of the project which can be used with an API by the Scrum team.  Another challenge is the integration of external services like penetration testing into the development process. One way to do so is that external resources provide test cases (e.g., for Metasploit \cite{linkmeta}) that can be used for every branch of the emerging software at any time. Results of tests can be documented as artifacts in the Backlog. Then they can be handled like any other change request.

Providing external view: One major part in \itsec is to recognize ways to exploit the own system. In other words, one must think like an attacker to recognize potential attack vectors.
Usually, it is easier for an outsider to spot potential weaknesses of a system than it is for the developer of a system. Hence, external resources may introduce a valuable external viewpoint on a project. When using the identification component of Secure Scrum, an external consultant can be helpful to point the team to security concerns. When using the implementation component, external resources can be helpful in the sprint planning. When using the verification component, an external consultant can help to create tests for security concerns. These interventions by external resources should not be part of the normal Scrum processes, the external resource should only help to ask questions (in the meaning of: he should show relevant concerns in scope of \itsecn). In conclusion, the external resource should help to set focus on problems the team is not aware of.
 
\section{Evaluation}\label{eval}
The evaluation presented in this paper focuses on the following questions: is Secure scrum a practicable approach to develop secure software? Is Secure Scrum easy to understand? Does Secure Scrum increase the security level of the developed software?

As test setting, 16 developers were asked to develop a small piece of software. The developers were third year students in computer sciences and business informatics (BSc). They were not aware that they are part of this evaluation. The students showed programming skills that were on the usual level of a third year bachelor student. No participant attended a specialized course in \itsec before beside the compulsory lecture in \itsec (basic level) in the second year of the bachelor. All developers had average theoretical knowledge about Scrum. Only two students had practical experiences (less than 2 months) with Scrum. No one had practical experiences in \itsecn. 

The developers were divided into three groups.
\begin{enumerate}
\item Team 1 (T1): The Anarchist group: They could manage themselves as they like, except using Scrum
\item Team 2 (T2): The Scrum group: They should use standard Scrum
\item Team 3 (T3):The Secure Scrum group:  They should use Secure Scrum.
\end{enumerate}

To avoid influences on the evaluation, teams $1$ and $2$ thought that team $3$ also uses standard Scrum. All groups got a list of six requirements for a new software product. They were asked to develop a prototype for a social network with the following features: registration, login, logout, personal messages, wall messages, bans, friend lists, and further more. Each group had only one week to develop this prototype using Java and a preconfigured spring framework template (based on BREW \cite{pohl2014}). Each group was asked to develop a piece of software including as many requirements as possible (they knew that it was impossible to implement all requirements for the final version of the software considering the harsh time constraints). They were also told that they need to \flex{sell} their prototype on the last day of the experiment in front of a jury. In fact they should learn how to present their prototype and act like a team that wants to have a contract for further development. This should ensure that every team needs to define for itself the selling points of their prototype.

Team 3 has a short briefing of about one hour about Secure Scrum. Every team is advised to make a proper documentation. This includes all produced artifacts, the sources, and a short description of their development process.

Table  \ref{tab:metric} summarizes some basic findings of the experiment.

\begin{table}[!t]
\renewcommand{\arraystretch}{1.3}
\caption{Results of the evaluation of the efficiency and effectiveness of Secure Scrum}
\label{tab:metric}
\centering
\begin{tabular}{l l | l l l}
\hline
\bfseries \# & \bfseries Metric & \bfseries T1& \bfseries T2 & \bfseries T3\\
\hline\hline
1 & Lines of Code & 1149 & 758 & 458 \\
2 & Number of Basic Requirements & 6 &  6 & 6\\
3 & Number of additional Requirements defined & 0 & 1 & 8\\
4 & Number of Basic Requirements documented &0 &6 & 6\\
5 & Number of Basic Requirements implemented & 6 & 5 & 4 \\
6 & Number of Requirements documented & 0 & 7 & 14\\
7 & Number of Requirements implemented & 6 & 6 & 9\\
8 & Number of Vulnerabilities $sp$ & 18 & 12 & 3\\
9 & Group size & 6  & 5  & 5\\
\end{tabular}
\end{table}

 All three teams had a rough definition of the six basic requirements which should be implemented. They were told that whenever the requirements list should be enhanced to deal with the 6 requirements given by the customer, they are free to define new requirements. Team 1 did not define any new requirements. Team 2 defined one new requirement to enhance performance. Team 3 defined 8 new requirements that had a focus on \itsecn. These requirements are an excerpt of the descriptions for the \stagn. Overall, they defined 29 new stories focused on \itsecn.
This shows that even with beginner skills in computer sciences and low skills in \itsecn, it is possible to define a high amount (compared to the original requirements) of security related requirements. It also shows that it is possible to describe the most problematic vulnerabilities or problems with the help of risk identification.

Metrics $4-7$ of table \ref{tab:metric} are used to evaluate if the teams documented all requirements and how many of the requirements were implemented. 
This shows that the teams did not take care of any further requirements when not specified by the customer. This sounds trivial, but it also shows that the developer did not take care of \itsec when not specified. The Secure Scrum team (team 3) is the only team that did not implement all basic requirements. Instead, they obviously prioritized some of the security requirements over the basic requirements as some of the additional requirements were implemented. This finding shows that Secure Scrum helps to put focus on software security.

Metric 8 shows the number of security problems that were created by the developers. The number of security problem is calculated as follows: Let $sl$ be a vulnerability listed in the OWASP Top 10 list $OTT$ ($sl \in OTT$) . The OWASP Top 10 project lists the most common security vulnerabilities for web applications, e.g., Injection, Broken Authentication and Session Management, Cross-Site Scripting (XSS), Insecure Direct Object References, Security Misconfiguration, Sensitive Data Exposure, Missing Function Level Access Control, Cross-Site Request Forgery, Using Components with Known Vulnerabilities, and Unvalidated Redirects and Forwards. Let $OS$ be the complete source code of the developed software and $SC$  the part of the software written by the students ( $SC \subset OS$ ). 
Let $cf$ be a Java function. Let $cpf(sl)$ be a function that counts the amount of $sl$ for one $cf$.
By definition, $cpf(sl)$ increments a vulnerability counter by one whenever the current function is the source $ms$ function for a vulnerability.
A function $cf$ is considered as a source $ms$ whenever $cf \in SC$ and when the function is the reason for the vulnerability or it calls a function $cf_1$ where $cf_1 \notin SC$ and $cf_1$ is the reason for the flaw. The amount of vulnerabilities $sp$ is the sum of all $cpf(cf)$. 
Such a definition of the number of security problems only counts code that is responsible for vulnerabilities of a software system. It also takes into consideration the use of flawed code. For example, when a developer creates an SQL statement with a potential SQL Injection flaw, the function holding the database call with this statement is regarded as the reason of the vulnerability. 
The results of the evaluation shows, that team 1 and team 2 had a high amount of vulnerabilities in their software (team 1: 18, team 2: 12). Both teams built software exploitable by SQL Injection, XSS, CSRF, and had a vulnerable session management.
Team 3 had significantly less vulnerabilities.  This shows, that the use of Secure Scrum increase the security level of the developed software.

The first metric (Lines of Code (LOC)) shows the amount of code which was generated during the week. There are significant differences between the three teams.  The teams that identified additional requirements (performance (team 2) and security (team 3)) were not as productive as the other teams. This shows the overhead that comes with a broadened focus on software quality, especially on non-functional requirements. 

To evaluate ease of use and practicability of Secure Scrum, the documentation of the Scrum teams was evaluated. 
The documentation consists of the Backlogs and a timetable.
Table \ref{tab:doc} summarizes the results of this evaluation.
\begin{table}[!t]
\renewcommand{\arraystretch}{1.3}
\caption{Results of the evaluation of the practicability of Secure Scrum}
\label{tab:doc}
\centering
\begin{tabular}{l l | l l l}
\hline
\bfseries \# & \bfseries Metric & \bfseries Team 2 & \bfseries Team 3\\
\hline\hline
1 & Number of  requirements & 7 & 14 \\
2 & Number of user stories & 7 (13) & 14 (62)\\
3 & Number of tasks & 18 & 35 \\
4 & Number of user stories with \mtag & - & 14\\
5 & Number of tasks with \mtag & - & 8(35)\\
\end{tabular}
\end{table}

The numbers in braces are the total amount of user stories. The aggregated number (not in braces) shows the amount of user stories when grouped together.
This means a group of user story is a ``bigger'' user story which reflects a requirement. Team 2 broke down every user story to a different task. 
Team 3 broke down tasks for only the stories that they also implemented. This is why they defined more user stories than tasks. 
Team 3 found for every user story some security concerns, this is why they tagged all user stories. Metric 5 shows that all tasks also had \mtagn s, overall they had 8 different groups in the tasks.
Team 3 decided to create the links by grouping, they simply used red cards for the descriptions to show security problems (\mtag). This also shows that the proposed tools are simple enough to adapt them very fast in a Scrum process.

In conclusion, the evaluation shows that Secure Scrum is able to improve the security level of the developed software. Secure Scrum is easy to understand, can be used in practice, and is even suitable for teams that have no deepened security knowledge. The evaluation also shows that it is possible to have a proper documentation through all stages of the experiment. The tools of Secure Scrum harmoniously blend  into the standard Scrum toolset without the need of much overhead for training. 

\section{Conclusion}\label{sec:conclusion}
This paper presents Secure Scrum, an extension of the software development framework Scrum. Secure Scrum enriches Scrum with features focusing on building secure software. One of the main contributions of Secure Scrum are \stagn s, a way to annotate Backlog items with security related information. Secure Scrum was evaluated in a small software development project. The evaluation shows that Secure Scrum can be used in practice, is easy to use and understand, and improves the level of software security. 
 
\bibliographystyle{IEEEtran}
\bibliography{bib.bib}

\end{document}